\documentclass[10pt,twocolumn]{article}
\usepackage{amssymb}
\usepackage{amsthm}
\usepackage{authblk}
\usepackage{setspace}
\usepackage[textwidth=6.5in,textheight=8.25in,centering]{geometry}
\usepackage[pdftex]{graphicx}

\newcommand{\cm}{cm$^{-1}$}
\newcommand{\ccm}{cm$^{-2}$}

\begin{document}

\title{Spectroscopic investigation of quantum confinement effects in ion implanted silicon-on-sapphire films}
\author{Rajesh Kumar\thanks{{\scriptsize Present Address: National Institute
for Nanotechnology (NINT), University of Alberta, Edmonton, Alberta,
Canada.}} \footnote{{\scriptsize Corresponding Author:
rajesh2@ualberta.ca}}, H.S. Mavi and A.K. Shukla} \affil{Department
of Physics, Indian Institute of Technology Delhi, New Delhi -
110016, India}
\date{}

\maketitle

\begin{abstract}
Crystalline Silicon-on-Sapphire (SOS) films were implanted with
boron (B$^+$) and phosphorous (P$^+$) ions. Different samples,
prepared by varying the ion dose in the range $10^{14}$ to 5 x
$10^{15}$ and ion energy in the range 150-350 keV, were investigated
by the Raman spectroscopy, photoluminescence (PL) spectroscopy and
glancing angle x-ray diffraction (GAXRD). The Raman results from
dose dependent B$^+$ implanted samples show red-shifted and
asymmetrically broadened Raman line-shape for B$^+$ dose greater
than $10^{14}$ ions cm$^{-2}$. The asymmetry and red shift in the
Raman line-shape is explained in terms of quantum confinement of
phonons in silicon nanostructures formed as a result of ion
implantation. PL spectra shows size dependent visible luminescence
at $\sim$ 1.9 eV at room temperature, which confirms the presence of
silicon nanostructures. Raman studies on P$^+$ implanted samples
were also done as a function of ion energy. The Raman results show
an amorphous top SOS surface for sample implanted with 150 keV P$^+$
ions of dose 5 x $10^{15}$ ions cm$^{-2}$. The nanostructures are
formed when the P$^+$ energy is increased to 350 keV by keeping the
ion dose fixed. The GAXRD results show consistency with the Raman
results.
\end{abstract}
\section{Introduction}
Semiconductor nanostructures (NSs) have been investigated in recent
years because of their immense use in electronic and opto-electronic
devices [1-3]. Silicon (Si) NSs can be prepared on a variety of
substrates (sapphire, quartz or semiconductors) for micro-electronic
photonic device fabrication. The Si NSs fabricated on sapphire
substrates have distinct advantages over crystalline Si (c-Si).
Si-on-sapphire (SOS) is used in large-scale integrated circuitry
primarily because it provides better electrical insulation to
prevent stray currents in the electronic circuits. It is also used
in aerospace and military applications because of its radiation
resistance properties [4, 5]. Apart from its technological
importance, Si NSs are also studied because of change in electronic,
vibrational and optical properties as a result of quantum
confinement effects.

The semiconductor structures can be fabricated using variety of
techniques, which include, rf co-sputtering [6, 7], plasma
decomposition of compounds [8, 9], laser-induced etching [10, 11]
and continuous wave laser annealing of amorphous Si film [12, 13].
Ion implantation followed by subsequent annealing is another widely
used technique for the fabrication of Si NSs [14-17]. Ion
implantation technique is used because of its simplicity and
compatibility with integrated circuits and of the ease of
controlling particle size and density of NSs. When Si implanted
SiO$_2$ films are annealed in nitrogen or vacuum, Si NSs are formed
giving rise to visible photoluminescence (PL) [18,19]. Generally in
ion implantation method, NSs of desired element can be fabricated by
implanting that particular ion into any matrix. Recently, Giri et
al. [20, 21] have shown that Si NSs can be fabricated by implanting
germanium ions into SiO$_2$ film on Si wafer followed by annealing
in an argon gas atmosphere. They showed that Si NSs are formed as a
result of radiation-induced transfer of energy to Si ions and their
nucleation and growth under suitable thermodynamic conditions.
Further investigations of the effects induced by ion-bombardment on
Si films must be done. Detailed studies can be done by varying the
dose and the implanted ion to look into possible Si NSs formation.
One may expect a damage of the top surface as a result of ion
implantation. Investigation of the damage level and possible
formation of Si NSs in ion implanted SOS films will be very useful
in integrating the SOS technology with microelectronics to make
optical devices like Si quantum dot lasers [22, 23].

Light scattering is one of the most powerful methods for studying
the structures of disordered or partially ordered solids [24, 25].
Quantitative distinction between a crystalline lattice and a lattice
where some degree of disorder has been introduced can be made by
Raman spectroscopy. Raman spectra of amorphous Si reveals a broad
band around 470 \cm [26], while the spectrum from crystalline form
shows a sharp peak at 520.5 \cm corresponding to the zone-centered
phonon [27]. On the other hand, zone-edge phonons, which appear only
in two-phonon Raman scattering, corresponds to large wave vectors
and are sensitive to short-range disorder [28]. In NSs of a few
nanometer size, a major modification is observed in vibrational
properties due to the confinement of phonons. Raman spectroscopy is
widely used technique for investigating the phonon confinement
effect in NSs [29-33]. The change in the Raman line-shape is useful
in determining the size, shape and size distribution of NSs, since
phonon softening and widening of Raman lines are related to the size
of NSs.

The purpose of this paper is to make a systematic and comparative
study of structural modifications in the SOS films implanted with
boron (B$^+$) and phosphorous (P$^+$) ions. Raman studies are done
on B implanted SOS films as a function of B$^+$ dose at constant ion
energy. Asymmetrically broadened and red shifted Raman line-shapes
are observed along with a hump at 475 \cm for samples implanted with
B$^+$ dose higher than 10$^{14}$ \ccm. The observed Raman
line-shapes are analyzed carefully using phonon confinement model
(PCM). It is found that there is a formation of Si NSs in the B$^+$
implanted samples along with an amorphous background. Raman analysis
was also done on SOS samples implanted with heavier (P$^+$) ions. In
P$^+$ implanted samples, the Si NSs are formed for higher ion energy
(350 keV) than that for B$^+$ implanted samples (150 keV) for same
ion dose. PL and glancing angle x-ray diffraction (GAXRD)
experiments are done to verify the Raman result. Furthermore, it is
observed that Si NSs can be formed on crystalline SOS films by
implanting an ion with an appropriate ion energy and ion dose
without any post annealing treatment.

\section{Experimental Details}
Thin crystalline SOS films of thickness 500 nm are grown by a
conventional CVD process. Samples B1, B2 and B3 are prepared from
crystalline SOS by implanting 150 keV B$^+$ ions with different
doses of 10$^{14}$, 10$^{15}$ and 5 x 10$^{15}$ ions \ccm
respectively. Samples P1 and P2 are prepared from crystalline SOS by
implanting P$^+$ ions with fixed fluence of 5 x 10$^{15}$ ions \ccm
and energy 150 keV and 350 keV respectively. Details of all the
studied samples are given in Table 1. The Raman and PL spectra were
recorded by employing a spectroscopic system consisting of a
SPEX-1403 doublemonochromator, a HAMAMATSU (R943-2) photomultiplier
tube and an argon ion laser (COHERENT, INNOVA 90). The Raman spectra
are recorded using an excitation photon energy $\sim$ 2.41 eV of the
argon-ion laser. The excitation photon energy of 2.6 eV was used for
PL recording at room temperature. Philips X$'$pert, pro PW 3040
x-ray diffractometer was used for x-ray studies in glancing angle
geometry with glancing angle of 2 degrees.\begin{table}
\caption{Sample details and summary of Raman results. Size of Si NSs
are estimated by fitting the experimental Raman data with the
theoretical line-shape using phonon confinement model, Eq. (1).}

\center {\tiny
\begin{tabular}{|c|c|c|c|c|c|}
  \hline 
  Sample & Ion & Ion energy & Fluence(cm$^{-2}$) & Raman peak &
  size
  \\ \hline
  B1 & B$^+$ & 150 keV& $10^{14}$ & 521cm$^{-1}$ & -- \\
  B2 & B$^+$ & 150 keV& $10^{15}$ & 518cm$^{-1}$ & 5 nm\\
  B3 & B$^+$ & 150 keV& $5\times10^{15}$ & 517cm$^{-1}$ & 4 nm\\
  P1 & P$^+$ & 150 keV& $5\times10^{15}$ & 475cm$^{-1}$ & --\\
  B3 & P$^+$ & 350 keV & $5\times10^{15}$ & 518cm$^{-1}$ & 5 nm\\
  \hline
\end{tabular}}

\end{table}
\section{Results and Discussion}
\subsection{Boron implanted samples}
Figure 1(a) shows the Raman spectrum of a crystalline SOS sample
showing a symmetric peak at 524 \cm, which is 3.5 \cm blue shifted
as compared to the c-Si (520.5 \cm). The phonon hardening of 3.5 \cm
indicates the presence of an interfacial compressive stress of 8.7 x
10$^9$ dyne \ccm according to Englert et al. [34]. Figures 1(b) -
(d) shows the Raman spectra from samples B1, B2 and B3 respectively.
The Raman peak positioned at 524 \cm from crystalline SOS shifts to
521 cm-1 when crystalline SOS is implanted with 150 keV B$^+$ ions
at a rate of 10$^{14}$ ions \ccm as shown in Fig. 1(b). The Raman
line-shape from sample B1 is symmetric with FWHM of 4.5 \cm.
\begin{figure}
\includegraphics[width=8.0cm]{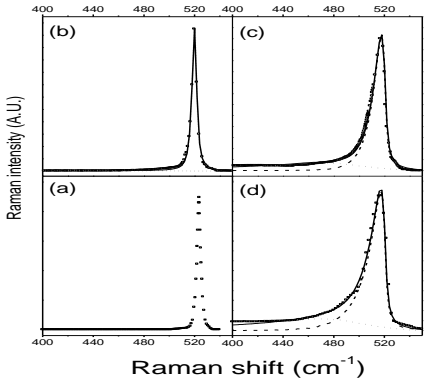}
\caption{Raman spectra from (a) crystalline SOS and B+ implanted
samples prepared by implanting 150 keV B$^+$ ions with a fluence
rate of (b) 10 ions \ccm, (c) 10$^{15}$ ions \ccm and (d) 5 x
10$^{15}$ ions \ccm. Discrete points are experimental data and solid
line is theoretically calculated Raman line-shape. Raman signals
from amorphous and nanocrystalline Si are shown by dotted and dashed
lines respectively.}
\end{figure}The
Raman peak shift from 524 \cm to 521 \cm may be because of
relaxation in the interfacial stress due to the B$^+$ ion
implantation. Because of ion implantation there is a little damage
to the film surface, which gives a path for the film-interface
stress to be relieved partially. This type of relaxation in the
stress has also been observed on porous Si on sapphire [35].
Dubbelday et al. [35] have shown that when the SOS film is stain
etched in HF: HNO$_3$ solution, crevices and cracks are formed in
the film. There is a relaxation in stress as a result of increase in
the porous Si layer thickness on the sapphire substrate. In our
sample B1, this type of relaxation in stress may take place through
the damage created by ion implantation. When the dose of B$^+$ ion
is increased to 10$^{15}$, the Raman peak position shifts to 518 \cm
and asymmetrically broadened with FWHM of 11.5 \cm as shown in Fig.
1(c). The Raman spectrum from sample B2 shows a hump around 475 \cm
indicating the presence of some amorphous Si background. On further
increasing the dose to 5 x 10$^{15}$ by keeping the ion energy same
(150 keV), the Raman peak downshifts further to 517 \cm with FWHM of
15.5 \cm. The amorphous component is also increased in sample B2
because of more amorphization due to increased B$^+$ dose.

\begin{figure}
\includegraphics[width=8.0cm]{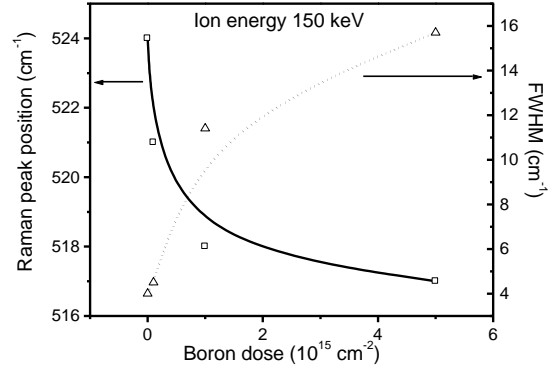}
\caption{Variation of Raman peak position and FWHM from B$^+$
implanted samples as a function of dose with fixed ion energy of 150
keV.}

\end{figure}
The variation of Raman peak position and FWHM as a function of B$^+$
implantation dose is shown in Fig. 2. From the Raman spectra of
B$^+$ ion implanted samples in Fig. 1, it can be seen that as the
dose is increased, there is an increase in the intensity of the
amorphous part (475 \cm). The amorphization of the top SOS surface
as a result of increased dose have been further verified by
recording the second order Raman scattering spectra as shown in Fig.
3. Figure 3(a)-(c) show the second order Raman spectra from samples
B1, B2 and B3 respectively. It is clear from Fig. 3 that the second
order Raman scattering intensity diminishes with increasing B$^+$
dose on the sample. This indicates the more and more amorphization
of the SOS films but complete amorphization has not taken place even
at the highest B+ dose of 5 x 10$^{15}$ ions \ccm in our case. A
gradual red shift is also observed in the Raman peak position in
Fig. 1. \begin{figure}
\includegraphics[width=8.0cm]{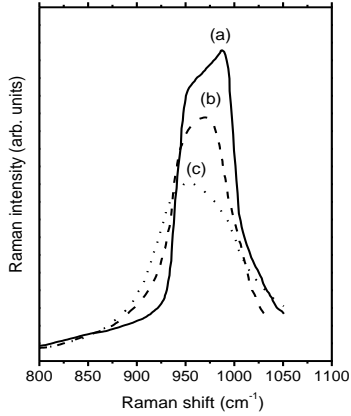}
\caption{Second order Raman spectra from (a) sample B1, (b) sample
B2 and (c) sample B3}
\end{figure}
To investigate the red shift the amorphous contribution has
been subtracted from the full Raman data in Figs. 1(c) and 1(d). The
amorphous part is shown by dotted line and the subtracted part is
shown by dashed line in Figs. 1(c) and 1(d). The Raman scattering
data obtained after subtraction of the amorphous part is red shifted
and asymmetric in nature. This type of behavior indicates the
presence of quantum confinement effect in the samples B2 and B3. The
quantum confinement of phonons is observed in the Si NSs, which have
been formed in the amorphous Si matrix because of ion implantation.
To get the information about the size of the Si NSs formed, the
subtracted Raman spectra in Figs. 1(c) and 1(d) were analyzed within
the framework of the PCM [36,37]. The PCM developed by Richter et
al. [36] and Campbell et al. [37] describes the Raman line-shape of
the optical phonons of low dimensional materials and has been widely
reported to estimate the NSs sizes [38, 39]. Taking the phonon
weighing function to be a Gaussian, the first-order Raman spectrum
can be expressed as:
\begin{equation}
I(\omega)\propto\int_{L_1}^{L_2}N(L)\left[\int_0^1\frac{e^{-\frac{q^2L^2}{4a^2}}}{[\omega-\omega(q)]^2+[\frac{\Gamma}{2}]^2}d^3q\right]dL
\end{equation}
where `q' is the phonon wave vector and  $\omega(q)$ is the phonon
dispersion relation for c-Si. The $\Gamma$ is the natural line width
of Raman mode and `L' is the confinement parameter (size of the
crystallite). The `L$_1$' and 'L$_2$' are the minimum and maximum
confinement dimensions respectively. The `N(L)' is a Gaussian
function of the form $N(L)\cong
e^{-\left(\frac{L-L_0}{\sigma}\right)^2}$ included to account for
the size distribution of the NSs. Where L$_0$ is the mean NSs size
and $\sigma$ defines the width of the distribution. The size of the
Si NSs obtained from the Raman line-shape fitting is displayed in
Table 1. The theoretical Raman line-shape given by Eq. (1) shows a
good fit with the experimental data in Figs. 1 (c) and 1(d). The
experimental data were best fitted with mean size of 5 nm for sample
B2 and with 4 nm for sample B3. The decrease in the NSs size with
increasing B$^+$ dose is also consistent with the red shift of the
Raman peak position.

\begin{figure}
\includegraphics[width=8.0cm]{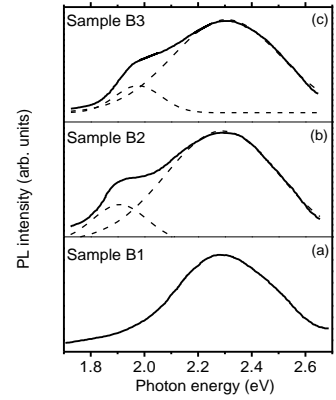}
\caption{PL spectra from (a) sample B1, (b) sample B2 and (c) sample
B3.}
\end{figure}

To further confirm the formation of Si NSs, PL spectroscopy is done
at room temperature. Figure 4(a)-(c) shows the PL spectra from
samples B1, B2 and B3 respectively. There are two important features
in Fig. 4. Firstly there is a broad PL peak centered around 2.28 eV,
which is independent of the B$^+$ ion dose. Secondly, there is an
appearance of a PL band at 1.9 eV, which  is absent in PL spectrum
from sample B1 in Fig. 4(a). The PL peak at 1.9 eV for sample B2
shifts to 1.95 eV when B+ dose is increased from 10$^{15}$   to 5 x
10$^{15}$ \ccm. Raman results in Table 1 show that the size of Si
NSs decrease with increasing the ion dose. Therefore, the size
dependent PL peak centered at 1.9-1.95 eV is attributed to the
quantum confinement of electrons in Si NSs[40]. This PL peak is
absent for the sample implanted with dose 10$^{14}$ \ccm in Fig.
1(a). On the other hand, the PL peak at 2.28 eV doesn't show any
dependence on the size (or implantation dose). This size independent
PL is attributed due to the transitions between two trapped defect
states as discussed by Wolkin et at. [41]. The PL results are in
consonance with Raman scattering results in Figs. 1(b)-(d). On the
basis of Raman and PL results from B$^+$ implanted films one can say
that quantum confined effects can be observed in ion implanted SOS
films.

\subsection{Phosphorous implanted samples}
\begin{figure}
\includegraphics[width=8.0cm]{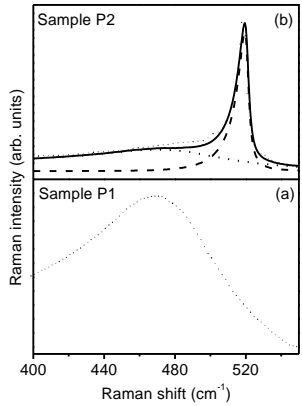}
\caption{Raman spectra from P+ ion implanted samples implanted with
dose 5 x 10$^{15}$ ions \ccm. The energy of P+ ions was 150 keV in
(a) and 350 keV in (b). Discrete points are experimental data and
the solid line is theoretically calculated Raman line-shape. Raman
signals from amorphous and nanocrystalline Si are shown by dotted
and dashed lines respectively in (b).}
\end{figure}To see the effect of implanted ion species on the crystalline SOS
samples, Raman studies have been done on P$^+$ ion implanted
samples. Figure 5(a) shows the Raman spectrum of P$^+$ ion implanted
sample (sample P1) with a dose of 5 x 10$^{15}$ and ion energy of
150 keV. One can see an amorphous structure around 475 \cm due to
implantation with a heavier ion as compared to B$^+$ implanted
samples. Figure 5(b) illustrates the Raman spectrum (hollow squares)
from sample P2. The Raman spectrum from sample P2 shows a sharp peak
at 518 \cm along with a hump at 475 \cm. The red shifted sharp peak
at 518 \cm indicates the presence of phonon confinement and needs
further analysis.

The Raman result from sample P2 has been further
analyzed in the same way as was done for the B$^+$ implanted Raman
results in the previous section. When the amorphous part is
subtracted from the full Raman data, an asymmetric Raman line-shape
is obtained as show by dashed line in Fig. 5(b). Subtracted data is
fitted with Eq. (1) to get the Si NSs size. We get the best fit of
experimental data for mean crystallite size of 5 nm which is less
than the Bohr exciton diameter $\sim$10 nm [42]. Thus we can say
that the asymmetry and red shift in the Raman line-shape in Fig.
5(b) is due to quantum confinement of phonons in the Si NSs formed
due to ion implantation. It can also be observed from Fig. 5(b) that
complete amorphization of the top SOS layer can be avoided by
increasing the ion energy. The ions with higher energy penetrate
deep into the film and the damage is less at the top surface. As a
result, partial amorphous Si along with Si NSs is formed.

\begin{figure}
\includegraphics[width=8.0cm]{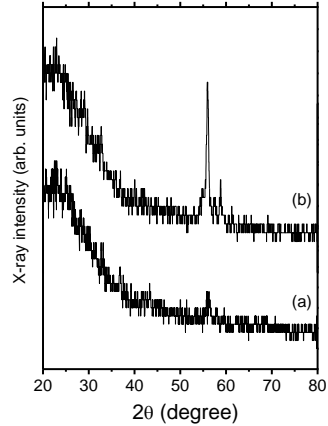}
\caption{X-ray diffraction spectra from (a) sample P1 and (b) sample
P2.}
\end{figure}
For further investigation of the top surface of the P$^+$ implanted
sample, GAXRD studies were done. Figures 6 (a)-(b) show the GAXRD
results of sample P1 and P2 respectively. The amorphous top surface
of sample P1 can be seen from Fig. 6(a). Whereas, for sample P2, the
GAXRD result shows a peak at 2$\theta \sim$ 56o corresponding to
(311) peak of Si in Fig. 6(b). Figure 6(b) suggests that the top
surface of sample P2 is not completely amorphous in nature. The
GAXRD results in Fig. 6 are in consonance with Raman scattering
results in Fig. 6. The Raman and GAXRD results confirm the presence
of crystalline Si phase in sample P2 and complete amorphous nature
of sample P1. By comparing the Raman results from B$^+$ and P$^+$
implanted samples one can say that when the crystalline SOS is
implanted with P$^+$ ions at the same dose and ion energy as of
B$^+$, complete amorphization of the top SOS layer take place. To
avoid the amorphization of the top layer, ions with higher energy
should be used for implantation. Under this condition, quantum
confinement from Si NS present in amorphous background can be seen
in Raman scattering. The Si NSs present in our samples can be formed
by implantation of any ion with appropriate dose and ion energy
without any post annealing treatment.

\section{Conclusions}
On the whole, a comparative Raman studies on B$^+$ and P$^+$ ion
implanted SOS samples are presented here.  It is found that the
inherent compressive stress at the interface in the crystalline SOS
film can be partially relieved when 150 keV B$^+$ ions were
bombarded with a fluence of 10$^{14}$ ions \ccm.  Raman line-shapes
show red-shift and asymmetry with increasing B$^+$ dose for a given
ion energy. An amorphous Si contribution also appears at B$^+$ dose
of 10$^{15}$ \ccm or higher. The top SOS surface is not completely
amorphous even at the B$^+$ dose of 5x10$^{15}$ \ccm when implanted
with B$^+$ ions. Phonon softening from 521 to 517 \cm and increase
in FWHM from 4.5 to 15.5 \cm are observed as a function of
increasing B$^+$ dose due to decrease in the size of Si NSs. The
true Raman line-shape because of the quantum confinement effect
alone is obtained by subtracting the amorphous contribution from the
full Raman data. The deconvoluted Raman data were analyzed within
the framework of PCM to obtain the size of Si NSs. Estimated value
of L$_0$ using Raman line-shape fitting varies between 4-5 nm for
B$^+$ implanted samples. The PL spectra of samples prepared by B$^+$
implantation on crystalline SOS show broad luminescence in the
visible region. The PL from higher dose samples show two peak
behavior. The 1.9 eV size dependent PL peak arises due to quantum
confinement of electrons in the Si NSs fabricated by ion
implantation. Whereas, the 2.28 eV PL peak is observed due to
defects in the sample.

Raman and GAXRD results show that when the crystalline SOS films are
implanted with heavier ions (P$^+$ in our case) by keeping other
parameters fixed, complete amorphization of the top SOS surface
takes place. The amorphization of top SOS layer can be avoided if
the energy of implantation is increased. The Si NCs of 5 nm sizes
are formed at higher ion energy in P+ implanted sample. The dose and
energy dependent Raman results from B$^+$ and P$^+$ implanted SOS
samples reveal Si quantum structures, are formed if complete
amorphization is avoided. The quantum confinement effect can be seen
only when the dose is higher than a critical value for a given ion
energy. The Si NSs in amorphous matrix can be formed by implantation
of any ion with appropriate dose and ion energy without any post
annealing treatment.

\begin{small}

\textbf{Acknowledgements}: The authors are grateful to Prof. V. D.
Vankar (IIT Delhi) for many useful discussions. The authors
acknowledge the financial support from the Department of Science and
Technology, Govt. of India under the project ``Linear and nonlinear
optical properties of semiconductor/metal nanoparticles for
optical/electronic devices". One of the author (R.K.) acknowledges
National Research Council (NRC-Nano project) for financial
assistance. Technical support from Mr. N.C. Nautiyal is also
acknowledged.
\end{small}

\end{document}